\documentclass[10pt,conference]{IEEEtran}

\usepackage{cite}

\usepackage{setspace}
\usepackage{epsf}\epsfverbosetrue
\usepackage{graphics,epsfig}
\usepackage{graphicx,epsfig}
\usepackage{epsfig}
\usepackage{multirow}
\usepackage{alltt}
\usepackage{subfigure}
\usepackage{tcolorbox}
\usepackage{float}
\usepackage{url}
\usepackage{graphicx}
\usepackage{verbatim} 
\usepackage{footnote} 
\usepackage{latexsym} 
\usepackage{sidecap} 
\usepackage{wrapfig}
\usepackage{eso-pic}
\usepackage{fix-cm}
\usepackage{algorithm}
\usepackage{algpseudocode}
\usepackage{color}
\usepackage{wrapfig}
\usepackage{multicol}
\usepackage{amsfonts} 

\usepackage{layout}

\usepackage{amsmath}
\usepackage[
  locale = DE 
]{siunitx}

\usepackage{textcomp}
\usepackage{multirow}
\usepackage{mathtools}
\usepackage{soul} 
 
\usepackage{verbatim}
\usepackage{breqn}
\usepackage{csvsimple} 
\usepackage{array}
\usepackage{subfig}
\usepackage[normalem]{ulem}
\usepackage{booktabs}
\newcolumntype{P}[1]{>{\RaggedRight\arraybackslash}p{#1}}

\usepackage{mathtools}

\usepackage{pifont}
\newtheorem{definition}{\textbf{Definition}}
\newtheorem{theorem}{\textbf{Theorem}}

\begin{document}


\title{Huff-DP: Huffman Coding based Differential Privacy Mechanism for Real-Time Data}

\author{\IEEEauthorblockN{Muneeb Ul Hassan\IEEEauthorrefmark{1}, Mubashir Husain Rehmani\IEEEauthorrefmark{4}, Jinjun Chen\IEEEauthorrefmark{1}\\}
\IEEEauthorblockA{\IEEEauthorrefmark{1}Swinburne University of Technology, Hawthorn VIC 3122, Australia\\ \IEEEauthorrefmark{4} Munster Technological University (MTU), Ireland}
}


\maketitle

\begin{abstract}
With the advancements in connected devices, a huge amount of real-time data is being generated. Efficient storage, transmission, and analysation of this real-time big data is important, as it serves a number of purposes ranging from decision making to fault prediction, etc. Alongside this, real-time big data has rigorous utility and privacy requirements, therefore, it is also significantly important to choose the handling strategies meticulously. One of the optimal way to store and transmit data in the form of lossless compression is Huffman coding, which compresses the data into a variable length binary stream. Similarly, in order to protect the privacy of such big data, differential privacy is being used nowadays, which perturbs the data on the basis of privacy budget and sensitivity. Nevertheless, traditional differential privacy mechanisms provide privacy guarantees. However, on the other hand, real-time data cannot be dealt as an ordinary set of records, because it usually has certain underlying patterns and cycles, which can be used for forming a link to a specific individuals’ private information that can lead to severe privacy leakages (e.g., analysing smart metering data can lead to classification of individuals daily routine). Thus, it is equally important to develop a privacy preservation model, which preserves the privacy on the basis of occurrences and patterns in the data. In this paper, we design a novel Huffman coding based differential privacy budget selection mechanism (\textbf{Huff-DP}), which selects the optimal privacy budget on the basis of privacy requirement for that specific record. In order to further enhance the budget determination, we propose static, sine, and fuzzy logic based decision algorithms. Furthermore, we carried out extensive theoretical analysis alongside performance evaluation and comparison on real-world datasets in order to show the effectiveness of our Huff-DP approach. From the experimental evaluations, it can be concluded that our proposed Huff-DP mechanism provides effective privacy protection alongside reducing the privacy budget computational cost.

\end{abstract}



\section{Introduction}

With the prompt development and adoption of emerging technologies (such as Internet of Things (IoT), next generation communication, blockchain, intelligent surfaces, mobile sensing, etc.), an exponential increase in the dependence upon interconnected devices has been seen~\cite{huffint01}. It has been reported by CISCO that the number of networked devices by 2023 will be 29.3 billion, as compared to 18.4 billion in 2018~\cite{huffint02}.  These networked devices generate a huge amount of data for a diverse range of applications, such as healthcare, smart homes, transportation \& travel, industries, education. Therefore, using this data can be very advantageous if utilized in an optimal manner. To handle this huge amount of data, research works are being carried out for its efficient storage, transmission, and analysation. On the other hand, if not handled properly, this data can lead to catastrophic consequences, because of the immense dependency on real-time everyday applications using such data~\cite{huffint03}.\\
In order to efficiently store, transmit, and use this real-time data, various compression and transmission models working over the functioning of channel coding, network coding, and source coding have been developed by researchers. However, considering the resource constrained nature of IoT, edge computing devices, and other networked devices, it is equally important to use computation and communication friendly mechanisms for storage and communication, which these devices can handle without bottlenecking the computational space~\cite{huffint04}. One such mechanism that is being used by researchers for resource constrained devices is Huffman coding (HC)~\cite{huffint05}. HC takes into account the redundant patterns/values and compress these values with the help of a binary tree in a lossless manner.\\
Since, the data is usually from real-time networked interconnected devices, thus, it is also important to know that this data cannot be dealt with as an ordinary set of records from a traditional data source. This is because of the reason that such fine-grained data usually have underlying patterns due to their almost-periodic nature. E.g., real-time data from smart metering nodes can be analysed to figure out the lifestyle pattern of residents (such as usage time of toaster, washing machine, etc.)~\cite{huffint06}. To protect privacy of real-time data, differential privacy is used as a de-facto standard. However, traditional differential privacy models treat all event instances as the same with a designated privacy budget. Contrarily, there could be rare events in the data that can leak more information as compared to usual events as per Shannon’s information measure property~\cite{huffint07}. E.g., the news with the title `it is snowing in the Sahara Desert’ contains much more information than the news `It is snowing at Mount Everest’ because of the reason that the first event is very rare. Similar is the case with real-time data, as the rare events can lead to leakage of more private information as compared to common events (e.g., a high usage of energy on a specific night can be linked to a party event in that smart home). Therefore, protecting privacy with respect to the rarity of events is also required alongside designing optimal compression.\\
In this paper, we first work over identification of required privacy nature of data with the help of Huffman coding and categorized data into various privacy levels. Afterwards, we work over choosing optimal differential privacy noise budget ($\varepsilon$) for each level in accordance with the required level of privacy. In order to choose optimal noisy budget ($\varepsilon$), we develop three decision mechanisms, named as static, sine, and fuzzy decision mechanisms. Collectively, we propose `\textbf{Huff-DP}’ mechanism, via which one can successfully protect the privacy of real-time data with respect to the underlying patterns in the data.

\subsection{Related Works}
A number of works protecting the privacy of real-time data with the help of various privacy protection models have been proposed so far. One such work via private preserving fuzzy counting from the perspective of streaming data has been carried out by~\cite{huffrel01}. In this work, authors proposed a notion of differential privacy that can be used to carry out count estimation in a private manner with the help of probabilistic data structures, such as Cuckoo and Bloom filter. Another very interesting work exploring the aspect of publishing a private dataset having time-interval characteristics with the help of differential privacy has been carried out by Jung~\textit{et al.} in~\cite{huffrel02}. Similarly, a work to quantify the privacy leakage alongside protecting it via differential privacy in a spatiotemporal manner for location data has been carried out by Cao~\textit{et al.} in~\cite{huffrel03}. Alongside quantifying the privacy leakage, this work also proposes a framework to transform the existing location privacy protection models into novel spatiotemporal event privacy protection mechanisms. Another work from the viewpoint of protecting the data privacy during continual release with the help of advanced local differential privacy settings have been carried out by Wang~\textit{et al.} in~\cite{huffrel04}. From the perspective of integration of privacy with lossless compression and Huffman coding, an interesting work in concealment of access of sequentially Huffman coded images for the authorized users has been carried out by authors in~\cite{huffrel05}. Moreover, the only work that discusses the computation and addition of noise based upon Huffman coding has been presented by Lin and Yang in~\cite{huffrel06}. The paper proposed a noise matrix and computes the distorted value from the specified noise matrix with the help of Huffman frequencies.\\
\noindent After analysing the related works, it can be concluded that to the best of our knowledge, no work in the literature discussed the integration and calculation of differentially private noise with the help of Huffman coding in real-time data settings. Similarly, no work proposed the notion of efficient differential privacy budget selection on the basis of occurrence probability of an event with the help of levels of Huffman tree.

\subsection{Key Contributions}

The key contributions of our Huff-DP work are as follows:

\begin{itemize}

\item We introduce the notion of Huffman coding based privacy preservation for real-time almost periodic data.
\item We propose the concept of integration differential privacy based protection for Huffman coded data.
\item We propose an efficient privacy preserving budget selection algorithm for optimal differential privacy noise generation with the help of static, sine, and fuzzy logic based decisions.
\item We enhance the utility of real-time almost symmetrical data alongside preserving privacy on the basis of occurrence probability.

\end{itemize}

\subsection{Paper Organization}

\noindent The rest of the article is organized in the following manner: Section 2 covers the preliminaries, motivation, problem formulation, case study based system model, and adversary model. Section 3 provides discussion about functioning and algorithms of Huff-DP mechanism. Section 4 provides performance evaluation and comparison with respect to various datasets. Finally, Section 05 concludes the paper by providing summary and insights.

\section{Providing Privacy for Real-Time Data}

In this section, we first discuss the preliminaries of Huff-DP and afterwards, we provide a detailed analysis from perspective of motivation, problem formulation, system model, and adversary model. 

\subsection{Preliminaries of Huff-DP}
\subsubsection{Huffman Coding}

The algorithm of Huffman coding has a legendary status in the disciplines of computing, whenever there is a discussion about computation of prefix-free minimum redundancy codes~\cite{huffpre01}. Nevertheless, Huffman is one of the most commonly used algorithmic techniques that has been applied to a diverse range of applications ranging from everyday communication to complex networks. Huffman coding, introduced by David Huffman in 1952 is an algorithm which can be used to construct minimum-redundancy codes for a given data on the basis of its probabilities~\cite{huffpre02}. In its true essence, Huffman coding is a greedy approach, which computes the minimum codeword length for a given set of symbols/data with respect to weight associated with that. Huffman coding solved the problem of computation of optimal binary prefix value of code ($C_f$) for the given distribution of frequency ($f(d)$) with respect to the length of each codeword ($L(C_w(d))$), which can be represented theoretically as follows~\cite{huffpre01}:

\begin{equation}
V(C_f) = \sum_{d = 1}^N f(d). L(C_w(d)) 
\end{equation}

In the above equation, $ V(C_f)$ is the sum selected codewords with respect to the frequency of occurrence of `$N$’ symbols. Thus, Huffman proposed an optimal mechanism to compute the minimum redundancy code in such a setting. A detailed discussion about functioning and implementation of Huffman coding is out of scope of this article. Interested readers are suggested to study a very good survey over Huffman coding by Alistair Moffat~\cite{ huffpre01}.

\subsubsection{Differential Privacy}

The term differential privacy was first pioneered in 2006 by Cynthia Dwork as a notion to preserve privacy of a specific individual in statistical databases by introducing a random noise in query results~\cite{DPRef01}. The notion works over the phenomenon that releasing a specific information should not reveal sufficient information that should be able to lead to identification of a single record in the data.~\cite{huffpre03}. A formal definitions of differential privacy protection~\cite{DPRef01} can be written as follows:

\begin{definition}\textbf{($\varepsilon$–Differential Privacy)}: A randomised computing mechanism $\mathbb{M}$ provides \textit{$\varepsilon$ – differential privacy} for selected neighbouring database pair $\mathcal{X}_1$ and $\mathcal{X}_2$ differing in a single element, having outcome set as $\mathcal{O}_p \subset Range(\mathbb{M})$ satisfies~\cite{huffint08}:

\begin{equation}
\label{EQ01}
\frac{Pr[\mathbb{M}(\mathcal{X}_1) \in \mathcal{O}_p]}{Pr[\mathbb{M}(\mathcal{X}_2) \in \mathcal{O}_p]} \leq  \exp(\varepsilon)
\end{equation}

\end{definition}

\begin{definition} \textbf{(Global Sensitivity)}: For any specified function $\mathcal{F}$: $\mathcal{X}_1 \rightarrow R^d$, the global sensitivity from the perspective of $\mathcal{X}_1$ can be defined as~\cite{huffint08}:

\begin{equation}
\Delta \mathcal{F} = \max_{\mathcal{X}_1,\mathcal{X}_2} ||\mathcal{F}(\mathcal{X}_1) - \mathcal{F}(\mathcal{X}_2)||
\end{equation}

\end{definition}
In this article, we work over optimizing the selection of $\varepsilon$ value in Eq.\ref{EQ01} via Huffman coding based privacy levels in order to enhance utility and privacy protection for probabilistic data.

\subsubsection{Fuzzy Logic}
Fuzzy logic works over the phenomenon of uncertainty in a decision with respect to imprecisely perceived information. E.g., instead of a traditional Boolean logic of 0 \& 1, the fuzzy logic can be in between, such as 0.7. In order to understand this better, one can take an example of a braking system in a car, which basically applies brakes in accordance with the distance with the front car. In case, if it is implemented with Boolean logic, then we will see sudden acceleration and sudden brakes. However, with the help of fuzzy logic, we can categorize the impact/intensity of brake and acceleration, such as 0.1, 0.2, etc. A detailed discussion about the fuzzy logic is out of scope of this article, interested readers can study~\cite{fuzzyref01, fuzzyref02, fuzzyref03}. In our Huff-DP scenario, we used this uncertainty and modelled the privacy budget selection with respect to the fuzzy operation. A detailed description of the integration of fuzzy logic in our Huff-DP model has been provided in Section.~\ref{FuzzySection}.

\subsection{Motivation for Huffman Coding based Differential Privacy}

Huffman coding is a widely adapted compression mechanism, which is being used by researchers and industry to efficiently compress the  probabilistic data into optimal shortened code for the purpose of storage and communication efficiency. Specifically, in the case of networked devices with certain resource constraints, Huffman coding can play a vital node by reducing the number of bits being used for transmission, storage, and processing of data~\cite{huffint05}. On the other hand, it is equally important to protect the privacy of such networked devices during communication, storage, processing, and analysis of data collected from them. In order to do so, various privacy preservation standards have been proposed, but currently differential privacy is being considered as a de-facto standard of privacy because of its strong theoretical guarantees~\cite{huffpre04}. Considering the aspect of integration of Huffman coding and differential privacy with real-time data from networked and other similar resources, we work over proposing an effective mechanism, which takes into account the benefits of both and complement each other in an optimal manner. Therefore, we propose \textit{Huff-DP} mechanism, which takes advantage of both, differential privacy, and Huffman coding. For instance, it utilizes the optimal compression of Huffman coding, and from the compression levels it generates an ideal privacy budget for efficient noise addition in the data.

\subsection{Problem Formulation \& Privacy Requirement in Real-Time Almost Periodic Data}

We divide this section into two parts: in the first part, we discuss the requirements of privacy for real-time almost periodic data and in the second part we demonstrate critical research questions that we addressed in the article.

\subsubsection{Privacy Requirements for almost Periodic Real-Time Data}

Data collected from real-time devices and scenarios cannot be termed as regular statistical data because it has certain underlying patterns in it which repeat themselves. E.g., charging patterns of a smart electric vehicle (EV) can be analysed to figure out the work and home routine of its owner, which can further be used to carry out malicious activities~\cite{huffpre05}. Similarly, web browsing patterns of a mobile/laptop user can be analysed to identify the mood or feelings of its user, this can further be used by recommender systems to carry out targeted advertisements~\cite{huffpre06}. Similarly, these sorts of patterns can be seen in almost every sort of real-time data, which can lead to leakage of users’ privacy. On top of that, this type of real-time data does also have certain rare occurrences, which can leak significantly more information as compared to common occurrences. E.g., a significantly high usage rate of smart meters can be linked to the happening of a gathering/party at a particular home~\cite{huffpre07}. Therefore, privacy preservation models also need to incorporate the notion of rare events while preserving privacy as a whole. \\
Generally, traditional differential privacy models are used to protect privacy in real-time scenarios, because differential privacy provides strong quantifiable theoretical guarantees for privacy. Nevertheless, traditional differential privacy models provide strong guarantees in conventional data settings, but in case of occurrence based privacy, traditional differential needs further enhancements. Similarly, in case of common events, the requirement for privacy is less as compared to a rare event. Thus, there is a need to design such differential privacy models, which calculate noise in accordance with the nature of the scenario. In this paper, we work over proposing a dynamic differential privacy budget selection model, which chooses privacy budget ($\varepsilon$) on the basis of its occurrence.

\subsubsection{Problem Questions}

The problems statement of our proposed mechanism can be classified into certain critical questions mentioned as below:

\begin{itemize}

\item How to integrate the notions of Huffman coding and differential privacy in real-time setting, so that they complement each other’s benefits?
\item How to classify the privacy requirements of real-time data with help of Huffman coding and event occurrence?
\item How to design a privacy preserving mechanism which preserves privacy of given data with respect to the required privacy?
\item How to dynamically allocate a differential privacy budget to every individual record on the basis of required noise?
\end{itemize}

\begin{figure}[t]        
\centering
\includegraphics[scale = 1.1]{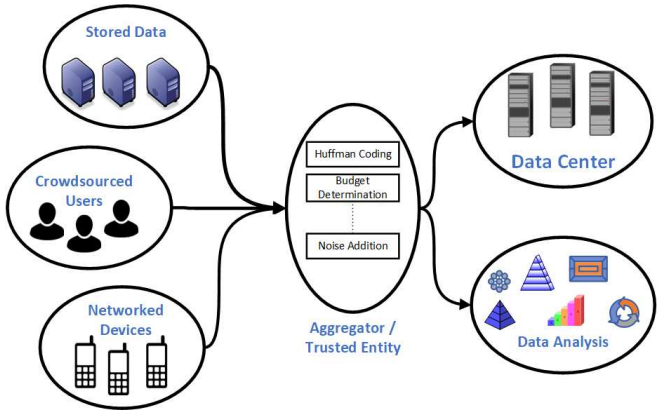}
 \caption{Graphical Illustration of System Model for Huff-DP Mechanism.}
  \label{fig:sysmodel}   
\end{figure}

\subsection{System Model}

Entities in our mechanism can vary dynamically depending upon the application and system parameters. However, in order to demonstrate the functioning of our proposed Huff-DP mechanism from a systematic perspective we made a generic system model, which has been presented in Fig.~\ref{fig:sysmodel}. By analysing the graphical illustration, the system model can be divided into three sub-types. One end of the system model comprises of data sources, such as cloud data centres, real-time users (e.g., smart meters), or networked devices (e.g., WiFi router). These entities transmit the fine-grained data to a centralized server/computing node in order to carry out privacy preserving operations to preserve privacy. On the other end of the figure are the data curators, such as data analysts, decision makers, or long-term storage data centres. The major purpose of these entities is to collect data given by centralized aggregator and take decisions, carry out various statistical analysis, or store them according to the nature of application. \\
In the middle of these entities lies a centralized server, which could be any trusted computational resource depending upon the nature of application. E.g., for cloud computing, it could  be an edge computing node, for a smart grid system, it could be a virtual power plant which is handling energy information from  a specified area, etc. The major purpose of this centralized aggregator is collection of data from sources in order to form a Huffman tree to compute optimal privacy budget via decision mechanism (static, fuzzy, or sine), which is further used to calculate the amount of added noise in the data. Finally, the computed noise is added by the aggregator and the final output is sent to data curators to carry out designated tasks.

\begin{figure}[t]        
\centering
\includegraphics[scale = 1]{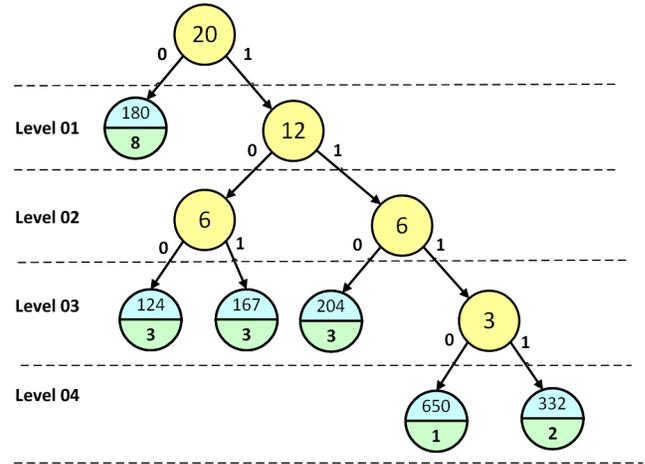}
 \caption{Example of Huffman Coding based Privacy Level Determination.}
  \label{fig:huffimg}   
\end{figure}

\begin{table}[H]
\begin{center}
 \centering
 \scriptsize
 \captionsetup{labelsep=space}
 \captionsetup{justification=centering}
 \caption{Tabular Demonstration of Huffman Table and Privacy Level.}
  \label{tab:compare}

\begin{tabular}{|p{5em}|p{2.7em}|p{2.7em}|p{2.7em}|p{2.7em}|p{2.7em}|p{2.7em}|}

\hline

 \centering \textbf{Value} & \textbf{180} & \textbf{124} & \textbf{167} & \textbf{204} & \textbf{332} & \textbf{650} \\ \hline

 \centering \textbf{Frequency} & 8 & 3 & 3 & 3 & 2 & 1 \\ \hline

 \centering \textbf{Code} & 0 & 100 & 101 & 110 & 1111 & 1110 \\ \hline

 \centering \textbf{Level} & 1 & 3 & 3 & 3 & 4 & 4 \\ \hline

 \centering \textbf{Required Privacy Level} & Low & Medium & Medium & Medium & High & High \\ \hline

 \end{tabular}
  \end{center}
\end{table}

\subsubsection{Example: Real-Time Data Release \& Privacy Leakage}
In order to demonstrate the functioning of our proposed Huff-DP model in a generic way, we develop a Huffman-tree from 20 arbitrary values of a smart meter. A graphical illustration of the designed tree can be visualized in Fig.~\ref{fig:huffimg}. In the figure, we took 20 arbitrary values of a smart meter with frequency distribution $<180:8,~124:3,~167:3,~204:3,~332:2,~650:1>$. It can be seen that the value `180’ has the highest frequency, therefore, it can be presumed that mixed usage multiple appliances can constitute this value (e.g., fan + monitor, or monitor + fridge, etc.). However, as we go down the spectrum, one can notice that the possibility of combinations reduces because of reduction in occurrence probability. E.g., the number of combinations of appliances to form an accumulative value  of `204’ is less than `180’. Therefore, this level needs more privacy as compared to the first level. Moving further to the bottom of the tree, it can be seen that `650 \& 332’ are the rarest occurrences, which can only be formed because of some very specific appliances. So, for a non-intrusive load monitoring (NILM) attacker or an adversary, it is very easy to identify the combination because of its rarity. Therefore, these rare events require more privacy protection as compared to common events. \\
Thus, in our proposed Huff-DP model, we take `node-depth’ into account while determining privacy budget and choose the differentially private noise with respect to the requirement of the situation. The same scenario can be applied to any other real-time/networked device, which has underlying patterns in its data that can lead to information privacy leakage. \\
It is also important to mention that Huffman-tree can be built in multiple alternative manners as well, e.g., by choosing the higher value on the right node or by starting the least frequency from the bottom right side, etc. However, it should be noted that the privacy levels of the tree usually do not get affected by choosing alternative methods, because they are determined on the basis of node depth. An explanatory table showing the values, frequency, chosen code, node depth, and privacy requirement has been given in Table.~\ref{tab:compare}.

\begin{algorithm}[t]\small
\caption{Huffman Coding Algorithm}
\label{HCAlgo}
\begin{algorithmic}[1]

\State $\textbf{Input} \gets F_q, N_o$ 
\State $\textbf{Output} \gets H_{t_n}$

\item[] $\textbf{FUNCTION} \rightarrow$ Huffman\_Coding$(F_q, N_o, H)$

\item[]//Arrange values nodes w.r.t increasing frequency.
\item[]//Initialize \& Create leaf nodes w.r.t increasing frequency.

\State \textbf{Set} $H$ $gets$ []

\For{(Every ($V_l$) in Freq distribution ($V_l \in $ [0,1,....($N_o$-1)]))}

	\State \textbf{Set} leaf-node $\gets$ new(leaf)
	\State \textbf{Set} leaf-node.val $\gets$ $V_l$
	\State \textbf{Set} leaf-node.freq $\gets$ $F_q(V_l)$
	
	\State \textbf{Insert} ($H$, leaf-node)

\EndFor

\While{$\left|H\right|$ $\geq$ 1}
\State \textbf{Set} leaf-node$_0$ $\gets$ Min\_Extract($H$)
\State \textbf{Set} leaf-node$_1$ $\gets$ Min\_Extract($H$)
\State \textbf{Set} leaf-node $\gets$ new(Internal\_Node)

\State \textbf{Set} leaf-node$_{lft}$ $\gets$ leaf-node$_0$ 
\State \textbf{Set} leaf-node$_{rgt}$ $\gets$ leaf-node$_1$ 

\State \textbf{Set} leaf-node$_{frq}$ $\gets$ leaf-node$_0$ +  leaf-node$_1$

\State \textbf{Insert} ($H$, leaf-node)
\EndWhile

\State \textbf{Set} leaf-node $\gets$ Min\_Extract($H$)
\State \textbf{return} leaf-node

\item[]//Huffman Tree is returned for further processing.

\end{algorithmic}
\end{algorithm}


\subsection{Adversary Model}
In the Huff-DP mechanism, the aim is to protect the identification of an individual record via differential privacy protection. The main adversaries in our model could be any intruder or curious analyst, who is intended to know more than the required information. E.g., the goal of collecting real-time values from smart meters should be to carry out load forecasting and other similar decision based statistical tasks, however, if instead of doing load-forecasting an analyst goes on to learn more about an individuals’ routine or about the health of appliances in a smart home, then this behaviour will be termed as adversarial behaviour. Similarly, in case of analysing the trip distances from airline passenger’s database, a non-adversarial will analyse the travelling habits of a city/country or any other similar statistics, however, contrarily, an adversarial analyst will try to identify the presence of a particular individual over a specific flight via linkage attacks. \\
Thus, we divide adversaries into multiple sub types; the first type could be adversaries, who are interested in knowing the private information for various non-threatening reasons, such as carrying out targeted advertisements, etc. The second type of adversary could be the entities, who are carrying out adversarial attacks because of various detrimental reasons, such as planning burglaries, etc. Irrespective of the nature of adversary, our aim via Huff-DP is to protect the private information of individuals from getting leaked.

\begin{algorithm}[t]\small
\caption{Privacy Budget Decision Algorithm}
\label{BDAlgo}
\begin{algorithmic}[1]

\State $\textbf{Input} \gets P_{L_r}, B_{d_f}$ 
\State $\textbf{Output} \gets \varepsilon_V$

\item[] $\textbf{FUNCTION} \rightarrow$ Budget\_Decision($P_{L_r}, B_{d_f},\varepsilon_V$)
\State \textbf{Initialize} Privacy Level ($P_{L_r}$), Budget Decision Function ($B_{d_f}$)

\If {$B_{d_f}$ == `Static'}
	\State \textbf{Initialize} $\beta_1$
	\State \textbf{Initialize} $S_p, F_p$ w.r.t Fig.~\ref{fig:decisionfunct}
	\State \textbf{Generate} $I_v$ = RAND($S_p,F_p$)
	\State \textbf{Calculate} $\varepsilon_{st}$ = $\beta_1 * I_v$
	
	\State \textbf{Set} $\varepsilon_V \gets \varepsilon_{st}$
\EndIf

\If {$B_{d_f}$ == `Sine'}
	\State \textbf{Initialize} $\beta_2$
	\State \textbf{Generate} $I_v$ = RAND($0,\pi$)
	\State \textbf{Calculate} $\varepsilon_{si}$ = $\beta_2 * (\frac{\sin(I_v)}{P_{LR}})$
	
	\State \textbf{Set} $\varepsilon_V \gets \varepsilon_{si}$
\EndIf

\If {$B_{d_f}$ == `Fuzzy'}
	\State \textbf{Initialize} $\beta_3$
	\State \textbf{Initialize} $S_{B_1},S_{B_2},F_{B_1}, F_{B_2}, C_{V_1}, C_{V_2}, b_p, c_p$ w.r.t Fig.~\ref{fig:decisionfunct}
	\State \textbf{Generate} $B_{V_1}$ = RAND($S_{B_1},S_{B_2}$)
	\State \textbf{Generate} $B_{V_2}$ = RAND($F_{B_1},F_{B_2}$)
	\State \textbf{Generate} $C_{V_t}$ = RAND($C_{V_1},C_{V_2}$)
	\item[] \;
	\State \textbf{Calculate} $\varepsilon_{fz}$ = $\beta_3 * (\frac{b_p*(B_{V_1} + B_{V_2}) + c_p*(C_{V_t})}{100})$
	
	\State \textbf{Set} $\varepsilon_V \gets \varepsilon_{fz}$
\EndIf

\State \textbf{return} $\varepsilon_V$

\item[]//Optimal Privacy Level $\varepsilon_V$ is returned for further processing.

\end{algorithmic}
\end{algorithm}


\begin{algorithm}[t]\small
\caption{Huffman Coding based Differential Privacy (Huff-DP) Algorithm}
\label{HuffDPAlgo}
\begin{algorithmic}[1]

\State $\textbf{Input} \gets V_d, B_{d_f}$ 
\State $\textbf{Output} \gets P_{R_v}$

\item[] $\textbf{FUNCTION} \rightarrow$ Huff\_DP$(V_d, B_{d_f},P_{R_v})$
\item[] \;
\item[]//Frequency Extraction and Huffman Coding Works as follows:

\State \textbf{Read} Real-Time Data ($V_d$)
\State \textbf{Extract} Frequency Distribution ($F_q \gets$ value \& repetition frequency)
\State \textbf{Set} Huff\_Tree = Huffman\_Coding($F_q$)
\item[] \;
\item[]//Budget Decision Algorithm works as follows:

\State \textbf{Initialize} Budget Decision Function ($B_{d_f}$)
\State \textbf{Initialize} Length of Huffman Tree ($L_h$)
\State \textbf{Initialize} First Node Depth ($F_{dn} = $ None)

\item[]//Calculating Depth of First Node:
\For{\texttt{(each \textbf{i} in \textbf{$L_h$})}}
\If {$F_{dn}$ is None or len(Huff\_Tree(i) $< F_{dn}$}
	\State $F_{dn} =$ len(Huff\_Tree(i))	
\EndIf
\EndFor

\item[]//Calculating Effective Node Depth:
\For{\texttt{(each \textbf{j} in \textbf{$L_h$})}}
\State $I_{dn} = $len(Huff\_Tree(j))
\State $E_{dn} = I_{dn} - (F_{dn} - 1)$ 
\State \textbf{Set} $P_{LR} \gets E_{dn}$ // For each value.
\EndFor

\item[]//Deciding Privacy Budget $\varepsilon$ for each value:
\For{\texttt{(each \textbf{j} in \textbf{$L_h$})}}
\State $\varepsilon_{V_j}$ = Budget\_Decision($P_{LR}, B_{d_f}$)$_j$
\EndFor

\item[] \;
\item[]//Calculating Noisy Values as follows:
\For{\texttt{(each \textbf{k} in \textbf{$L_h$})}}
\State \textbf{Initialize} Sensitivity ($\Delta \mathcal{F}$), mean ($\mu$)
\State \textbf{Set} Noise$_k$ = $N_v$ = $Lap(\mu, \varepsilon_{V_j}, \Delta \mathcal{F}$
\State \textbf{Set} Perturbed_Value$_k$ = $P_{R_v}$ = ABS($[V_{d_k} + N_v]$)
\State \textbf{Replace} $V_{d_k}$ w.r.t $P_{R_v}$
\State \textbf{Return} $P_{R_v}$
\EndFor

\item[]//$P_{R_v}$ is list of protected values for transmission, analysis, etc.

\end{algorithmic}
\end{algorithm}

\begin{figure*}[t]        
\centering
\includegraphics[scale = 1.5]{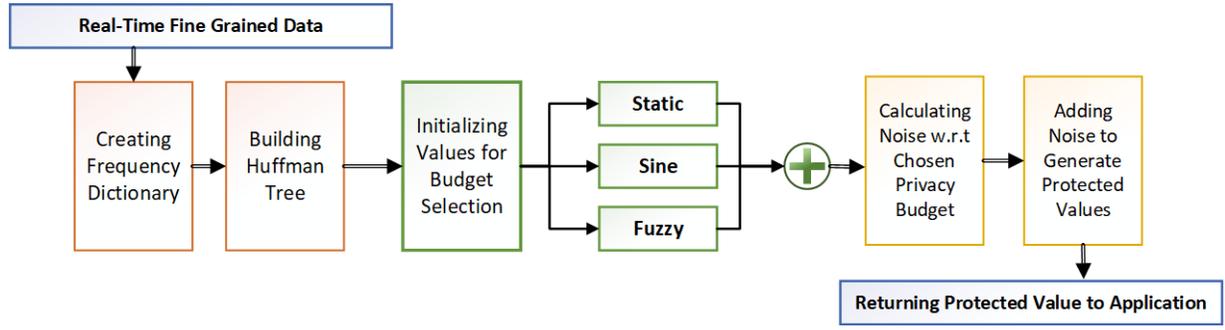}
 \caption{Example of Huffman Coding based Privacy Level Determination.}
  \label{fig:huffdp}   
\end{figure*}

\section{Huff-DP Mechanism and its Fundamental Components}

We divide the discussion about fundamentals of Huff-DP mechanism into two parts, firstly, we discuss the associated algorithms and afterwards, we provide discussion about theoretical and systematic analysis.

\subsection{Functioning of Huff-DP Mechanism (Huff-DP Algorithm)}

In this section, we provide a detailed analysis from the perspective of functioning of our Huff-DP algorithm. The discussion is divided into three sub parts, where firstly we discuss the formulation of Huffman code from the given data, after that, privacy level and budget determinations is carried out, and finally noise is computed on the basis of selected privacy level. Formulated algorithms have been given in Algo.~\ref{HCAlgo},~\ref{BDAlgo},~\ref{HuffDPAlgo}. Moreover, a graphical illustration of the proposed Huff-DP mechanism has been provided in Fig.~\ref{fig:huffdp}.

\subsubsection{Huffman Coding Part of Huff-DP}
The first part of Huff-DP algorithm comprises computation of Huffman coding tree in order to determine privacy levels. This part covers Line 2 to Line 5 of the Algo.~\ref{HuffDPAlgo}. Firstly, after getting the data input, the frequency of each valued real-time data ($V_d$) is computed and stored in the form of a frequency distribution. Afterwards, the `$Huffman_Coding$ function from Algo.~\ref{HCAlgo} is called, which firstly organize and creates empty leaf nodes with respect to occurrence frequencies of every character reading ($V_l$). After that, Huffman tree is created, where firstly the minimum value is extracted from the given distribution and appended as leaf-node$_0$. Similarly, the next value is appended as  leaf-node$_1$. After that, these leaf nodes are appended in the form of left and right branches respectively. Finally, the node is inserted to the distribution and these steps repeat till the formation of complete Huffman tree.\\
It is important to note that we followed a conventional way to build Huffman tree, and Huffman tree can be constructed in multiple ways, with respect to the arrangement of leaf nodes. Here, we demonstrate one of the most common methods. However, the assumptions and demonstration of Huff-DP mechanism also holds true, even if it is being built via any other method because the privacy budget decision depends upon the extracted code and node depth. A detailed demonstration of multiple methods of construction of Huffman is out of scope of this article. However, interested readers can study the article~\cite{huffpre01} to get deeper insight about other methods.

\begin{figure}[t]        
\centering
\includegraphics[scale = 0.72]{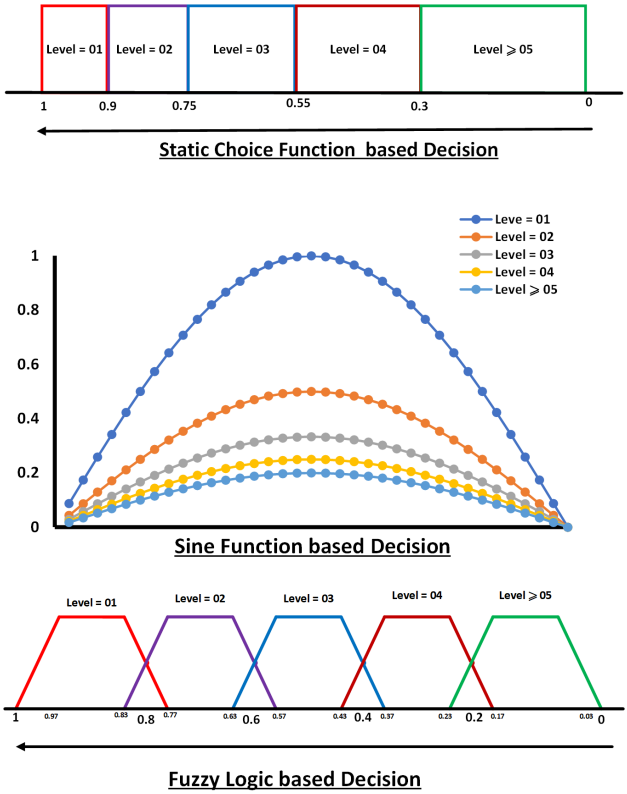}
 \caption{Illustrative Demonstration of Decision Functions used in Huff-DP Mechanism (Static, Sine, \& Fuzzy Logic).}
  \label{fig:decisionfunct}   
\end{figure}

\subsubsection{Privacy Level \& Budget Decision Part of Huff-DP}
After successful formation of Huffman tree, privacy level privacy budget is determined from Line 6 to Line 21. It is important to highlight that privacy level is determined on the basis of node depth of Huffman tree. In order to do so, first of all, the tree is iterated to figure out the depth of the first data node of the tree, which is stored as $F_{dn}$. After that, effective node depth $E_{dn}$ is computed with the help of $F_{dn}$ as follows:

\begin{equation}
E_{dn} = I_{dn} – (F_{dn} – 1)
\end{equation}

The value of $E_{dn}$ is further used to determine privacy level $P_{LR}$ according to the determined rules (given in Fig.~\ref{fig:decisionfunct}). Afterwards, the budget decision function $Budget_Decision()$ is called, where the values of $P_{LR}$ and $B_{df}$ are passed through to determine the optimal budget (Algo.~\ref{BDAlgo}). The budget decision algorithm comprises three sub-parts, and choice of specific part is done on the basis of $B_{df}$, which basically choses which specific model should be used to compute the privacy budget. The detailed description of these parts is given as follows:

\paragraph{Static Budget Selection}

The first and the simplest decision model in our Huff-DP strategy is static budget selection. In this decision, the budget values are determined with respect to a static random $\varepsilon$ generation with the help of uniform distribution from the perspective of privacy level. We formulate the decision graph into five sub choice sets having a different range of budget values (given in Fig.~\ref{fig:decisionfunct}). Each set is given a range of `0.2’, and the choice of a particular set is made on the basis of $P_{LR}$. For example, if the value of is 2, then the set range corresponding to level 2 is chosen, which comprises of $Range(0.6,0.8)$ in current case scenario. After determining the level, the values of $S_p$ (starting point) and $F_p$ (final point) is determined. Similarly, the hyper parameter $\beta_1$ is chose with respect to the application requirement. Finally, all these values are then used to compute final privacy budget with the help of following equation:

\begin{equation}
\label{staticbudget}
\varepsilon_{st} = \beta_1 * RAND(S_p, F_p)
\end{equation}

\paragraph{Sine Wave based Decision}

In order to introduce additional randomization in our decision model, we use the notion of sine function based selection. As we all are familiar that `sine’ function oscillates periodically with respect to the given input. E.g., the outcome of sine wave oscillates between 0 to 1 for the values between [0 - $\pi$] in a wave like pattern, and it shows exactly inverse behaviour ranging from [$\pi$- 0]. Thus, in order to determine the randomized value from sine function, we first restricted our output from [0 - 1] by giving a randomized input between 0 - $\pi$. Afterwards, the calculated value is divided by the node depth ($P_{LR}$). However, in order to control the excessive reduction of privacy budget, we restrict the node depth in the equation to a maximum of 5. Afterwards, the result is multiplied with a hyper parameter ($\beta_2$) in order to control the epsilon if required by application scenario. The equation to compute the privacy budget with respect to sine function is as follows:
\begin{equation}
\label{sinebudget}
\varepsilon_{si} = \beta_2 * \biggr[\frac{Sin(RAND(0,\pi))}{P_{LR}}\biggr]
\end{equation}

\paragraph{Fuzzy Logic based Selection} \label{FuzzySection}

The third and the most exhaustive decision making model in our Huff-DP mechanism is based upon the fuzzy logic phenomenon~\cite{huffalgo01}. Fuzzy logic can be considered as a methodology of computing, via which data can be modelled with the help of approximate modelling and reasonings. E.g., in our scenario, let us say that the chosen privacy level is 02, so there is a possibility that it may require some initial boundary parameters of level 02, or it may require some end boundary parameters, instead of a completely random value. Therefore, in order to model this approximate randomness in decision making, we introduced the notion of fuzzy logic based on $\varepsilon$ decisions. A graphical illustration of fuzzy logic can be seen in the third image in Fig.~\ref{fig:decisionfunct}. For example, in case of level 02 of privacy, starting boundary ranges from (0.57 – 0.63), similarly, ending boundary ranges from (0.77 – 9,83), while the core value ranges from (0.63 – 0.77). Thus, in our fuzzy Huff-DP budget selection model, we took the weightages as 20\% of each boundary value ($S_{B_1}, S_{B_2}, F_{B_1}, S_{F_1}$) and 60\% of core value (C_{V_1}, C_{V_1}. Similarly, hyper parameter ($\beta_3$) can be used to calculate the final value of epsilon, if required. The formal equation for computation of fuzzy logic based epsilon has been given as follows:

\begin{equation}
\nonumber
\resizebox{1\hsize}{!}{$ \varepsilon_{fz} = \beta_3 * \biggr[ \frac{ b_p *\bigr[ RAND(S_{B_1}, S_{B_2}) + RAND(F_{B_1}, F_{B_2})  \bigr] + c_p * \bigr[ RAND(C_{V_1}, C_{V_2})\bigr] }{100} \biggr]$}
\end{equation}

Considering $B_{V_1} = RAND(S_{B_1}, S_{B_2})$, $B_{V_2} = RAND(F_{B_1}, F_{B_2}$ and $C_{V_t} = RAND(C_{V_1}, C_{V_2})$, the above equation can be simplified as follows:

\begin{equation}
\label{fuzzybudget}
 \varepsilon_{fz} = \beta_3 * \biggr[ \frac{ b_p (B_{V_1} + B_{V_2})  + c_p * C_{V_t}}{100} \biggr]
\end{equation}

\noindent \textit{It is important to highlight that the values of all three decision models and involved hyper parameters can be changed as per the need of the scenario/application. We use the values in Fig.~\ref{fig:decisionfunct} in order to elaborate the functioning of our Huff-DP mechanism.}

\subsubsection{Noise Addition Part of Huff-DP}

The third and the final step in our Huff-DP algorithm is the calculation and addition of noise in the given data with respect to chosen $\varepsilon$ value. In this step, firstly the required values, such as sensitivity ($\Delta \mathcal(F)$, mean ($\mu$), etc. are initialized. Afterwards, the values are fed to the Laplace noise calculation function, which calculates the differentially private Laplacian noise as per the following formula~\cite{huffint09}:

\begin{equation}
\nonumber
\label{probeqn01}
\resizebox{0.65\hsize}{!} {$Lap(C_R, \mu, \Delta\mathcal{F}) = \frac{1}{2\frac{ \Delta\mathcal{F}}{\varepsilon_{val}}}. e^{\Bigg(-\frac{|C_r-\mu|}{\frac{\Delta\mathcal{F}}{\varepsilon_{val}}}\Bigg)}$}
\end{equation}

After calculation of noise, it is then added to the original character value ($C_r$). Afterwards, the noisy value is replaced with the original value for the whole Huffman tree distribution. Finally, the protected values ($P_{R_v}$) are returned to carry out required procedures, such as transmission, storage, analysis, etc.

\subsection{Theoretical \& Systematic Analysis of Huff-DP}

In this section, we provide analysis about theoretical and formulations of our Huff-DP model with respect to differential privacy guarantees and attacks resilience. 

\subsubsection{Differential Privacy Analysis of Huff-DP Mechanism}

The perturbation phenomenon in our Huff-DP mechanism works over differential privacy. Thus, in order to demonstrate that our noise addition and budget selection models are compliant with differential privacy guarantees, we provide statistical proofs for each budget selection model.  The evaluation with respect to the selection models and Huff-DP mechanism is given as follows:

\begin{theorem} \textit{The static budget selection mechanism of our proposed Huff-DP mechanism satisfies $\varepsilon_{st}$-differential privacy guarantees~\cite{DPRef02}}:
\label{statictheorem}
\hspace{10mm}\textit{\textbf{Proof:}} With the static privacy budget selection function, the value of $\varepsilon_{st}$ = $\beta_1*$RAND$(S_p,F_p)$ from Eq.~\ref{staticbudget} will be used. Let us consider $H_{dp}$ \& $H_{dp}^\prime$ $\in C^{|Y|}$ in a way that $|| H_{dp} - H_{dp}^\prime||_1 \leq 1$. The arbitrary string length up to $`k’$ for $H_{dp}$ \& $ H_{dp}^\prime$ will be $C_R = \{C_{R_1}, C_{R_2}, . . , C_{R_i}\}$. Thus, provided that both $ H_{dp}$ \& $ H_{dp}^\prime$ can further be connected with Laplace noise distribution with respect to probability density function as $P_{H_{dp}} \& P_{ H_{dp}^\prime}$ respectively. Thus, the two functions of probability can be matched at the given random string (according to Laplace theorem in~\cite{DPRef02}) as follows:


\[\noindent \frac{P_{H_{dp}} \left[C_R = \{C_{R_1}, C_{R_2}, . . , C_{R_i}\}\right]}
{P_{H_{dp}^\prime}\left[C_R = \{C_{R_1}, C_{R_2}, . . , C_{R_i}\}\right]}  = \nonumber \\\]
\[ ~~~~~~~~~~~ \prod_{j=1}^{k} 
\frac{\exp\left(- \frac{\beta_1*RAND(S_p,F_p) |\mathcal{F}(H_{dp})_k - C_{R_k}|}{\Delta \mathcal{F}}\right)}
{\exp\left(- \frac{\beta_1*RAND(S_p,F_p) |\mathcal{F}(M_{n}^\prime)_k - C_{R_k}|}{\Delta \mathcal{F}}\right)}\]

\begin{equation}
\nonumber
\resizebox{1\hsize}{!} {$= \prod_{k=1}^{N} \exp\left(\frac{\beta_1*RAND(S_p,F_p) ( |\mathcal{F}(H_{dp}^\prime)_k - C_{R_k}| - |\mathcal{F}(H_{dp})_k - C_{R_k}|)}{\Delta \mathcal{F}}\right)$}
\end{equation}

\begin{equation}
\nonumber
 \resizebox{1\hsize}{!} {$\leq \prod_{k=1}^{N} \exp\left(\frac{\beta_1*RAND(S_p,F_p) ( |\mathcal{F}(H_{dp})_k - |\mathcal{F}(H_{dp}^\prime)_k |)}{\Delta \mathcal{F}}\right)$}
\end{equation}

\[= \exp\left(\frac{\beta_1*RAND(S_p,F_p) ( ||\mathcal{F}(H_{dp}) - |\mathcal{F}(H_{dp}^\prime)||)}{\Delta \mathcal{F}}\right)\]
\[ \leq \exp (\beta_1*RAND(S_p,F_p))\]
\[\text{The value~}\beta_1*RAND(S_p,F_p)\text{~can be substituted as~}\]
\[\leq \exp (\varepsilon_{st}) \]


\end{theorem}

\begin{theorem} \textit{The sine function based budget selection mechanism of our proposed Huff-DP mechanism satisfies $\varepsilon_{si}$-differential privacy guarantees~\cite{DPRef02}}:
\label{sinetheorem}
\newline\hspace{10mm}\textit{\textbf{Proof:}} With the sine privacy budget selection function, the value of {\footnotesize $\varepsilon_{si}$ = $\beta_2 * \bigr[\frac{Sin(RAND(0,\pi))}{P_{LR}}\bigr]$}\normalsize from Eq.~\ref{sinebudget} will be used. Furthermore, taking the arbitrary values and neighbouring datasets with respect to Theorem.~\ref{statictheorem}, the functions can be written as:

\[\noindent \frac{P_{H_{dp}} \left[C_R = \{C_{R_1}, C_{R_2}, . . , C_{R_i}\}\right]}
{P_{H_{dp}^\prime}\left[C_R = \{C_{R_1}, C_{R_2}, . . , C_{R_i}\}\right]}  = \nonumber \]
\[ ~~~~~~~~~~~ \prod_{j=1}^{k} 
\frac{\exp\left(- \frac{\beta_2 * Sin(RAND(0,\pi)) |\mathcal{F}(H_{dp})_k - C_{R_k}|}{\Delta \mathcal{F}.P_{LR}}\right)}
{\exp\left(- \frac{\beta_2 * Sin(RAND(0,\pi)) |\mathcal{F}(M_{n}^\prime)_k - C_{R_k}|}{\Delta \mathcal{F}.P_{LR}}\right)}\]

\begin{equation}
\nonumber
\resizebox{1\hsize}{!} {$  = \prod_{k=1}^{N} \exp\left(\frac{\beta_2 * Sin(RAND(0,\pi)) ( |\mathcal{F}(H_{dp}^\prime)_k - C_{R_k}| - |\mathcal{F}(H_{dp})_k - C_{R_k}|)}{\Delta \mathcal{F}.P_{LR}}\right)$}
\end{equation}

\begin{equation}
\nonumber
 \resizebox{1\hsize}{!} {$\leq \prod_{k=1}^{N} \exp\left(\frac{\beta_2 * Sin(RAND(0,\pi)) ( |\mathcal{F}(H_{dp})_k - |\mathcal{F}(H_{dp}^\prime)_k |)}{\Delta \mathcal{F}.P_{LR}}\right)$}
\end{equation}

\[= \exp\left(\frac{\beta_2 * Sin(RAND(0,\pi)) ( ||\mathcal{F}(H_{dp}) - |\mathcal{F}(H_{dp}^\prime)||)}{\Delta \mathcal{F}.P_{LR}}\right)\]
\[ \leq \exp (\frac{\beta_2 * Sin(RAND(0,\pi))}{P_{LR}})\]
\[\text{:- The value~}\frac{\beta_2 * Sin(RAND(0,\pi))}{P_{LR}}\text{~can be substituted as~}\]
\[\leq \exp (\varepsilon_{si}) \]

\end{theorem}

\begin{theorem} \textit{The fuzzy function based budget selection mechanism of our proposed Huff-DP mechanism satisfies $\varepsilon_{fz}$-differential privacy guarantees~\cite{DPRef02}}:
\label{fuzzytheorem}
\newline\hspace{10mm}\textit{\textbf{Proof:}} With the fuzzy logic based privacy budget selection, the value of \footnotesize${\varepsilon_{fz} = \beta_3 * \bigr[ \frac{ b_p (B_{V_1} + B_{V_2})  + c_p * C_{V_t}}{100} \bigr]}$ \normalsize from Eq.~\ref{fuzzybudget} will be used. Furthermore, taking the arbitrary values and neighbouring datasets with respect to Theorem.~\ref{statictheorem}, the functions can be written as:


\[\noindent \frac{P_{H_{dp}} \left[C_R = \{C_{R_1}, C_{R_2}, . . , C_{R_i}\}\right]}
{P_{H_{dp}^\prime}\left[C_R = \{C_{R_1}, C_{R_2}, . . , C_{R_i}\}\right]}  = \nonumber \]
\[ ~~~~~~~~~~~ \prod_{j=1}^{k} 
\frac{\exp\left(- \frac{[\beta_3 * b_p (B_{V_1} + B_{V_2}) + (c_p * C_{V_t})] |\mathcal{F}(H_{dp})_k - C_{R_k}|}{100.\Delta \mathcal{F}}\right)}
{\exp\left(- \frac{[\beta_3 * b_p (B_{V_1} + B_{V_2}) + (c_p * C_{V_t})] |\mathcal{F}(M_{n}^\prime)_k - C_{R_k}|}{100.\Delta \mathcal{F}}\right)}\]

\begin{equation}
\nonumber
 \resizebox{1\hsize}{!} {$  = \prod_{k=1}^{N} \exp\left(\frac{[\beta_3 * b_p (B_{V_1} + B_{V_2}) + (c_p * C_{V_t}) ]( |\mathcal{F}(H_{dp}^\prime)_k - C_{R_k}| - |\mathcal{F}(H_{dp})_k - C_{R_k}|)}{100.\Delta \mathcal{F}}\right)$}
\end{equation}

\begin{equation}
\nonumber
 \resizebox{1\hsize}{!} {$\leq \prod_{k=1}^{N} \exp\left(\frac{[\beta_3 * b_p (B_{V_1} + B_{V_2}) + (c_p * C_{V_t})] ( |\mathcal{F}(H_{dp})_k - |\mathcal{F}(H_{dp}^\prime)_k |)}{100.\Delta \mathcal{F}}\right)$}
\end{equation}

\begin{equation}
\nonumber
 \resizebox{1\hsize}{!} {$= \exp\left(\frac{[\beta_3 * b_p (B_{V_1} + B_{V_2}) + (c_p * C_{V_t})] ( ||\mathcal{F}(H_{dp}) - |\mathcal{F}(H_{dp}^\prime)||)}{100.\Delta \mathcal{F}}\right)$}
\end{equation}

\[ \leq \exp (\frac{\beta_3 * b_p (B_{V_1} + B_{V_2}) + c_p * C_{V_t}}{100})\]

\begin{equation}
\nonumber
 \resizebox{1\hsize}{!} {$\text{:- The value~}\frac{\beta_3 * b_p (B_{V_1} + B_{V_2}) + c_p * C_{V_t}}{100}\text{~can be substituted as~}\varepsilon_{fz} \text{~as:}$}
\end{equation}

\[\leq \exp (\varepsilon_{fz}) \]


\end{theorem}




\begin{figure*}[t]
\centering
\captionsetup{justification=centering}
\begin{center}

\subfigure[]{
\includegraphics[scale = 0.33]{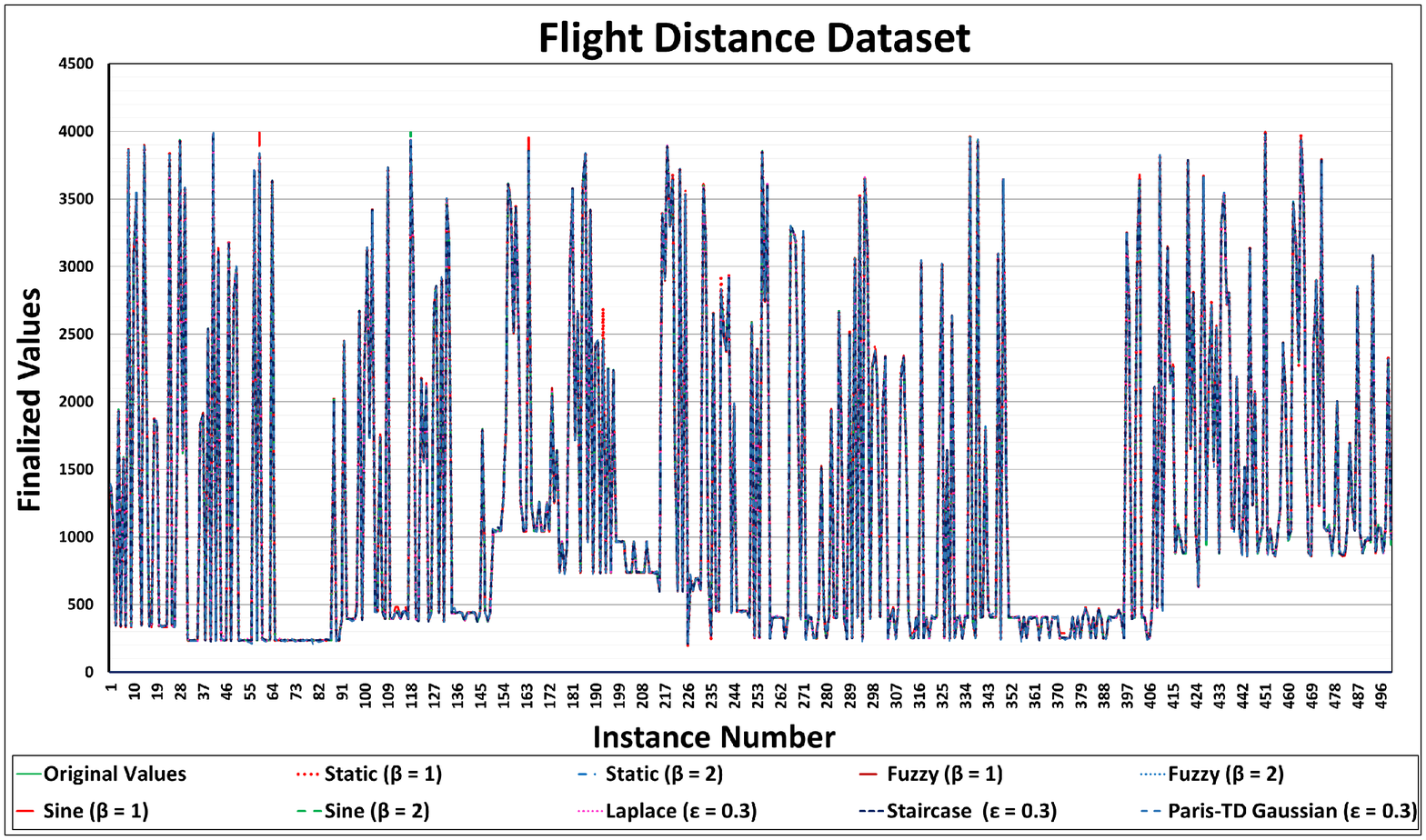}
}
\subfigure[]{
\includegraphics[scale = 0.33]{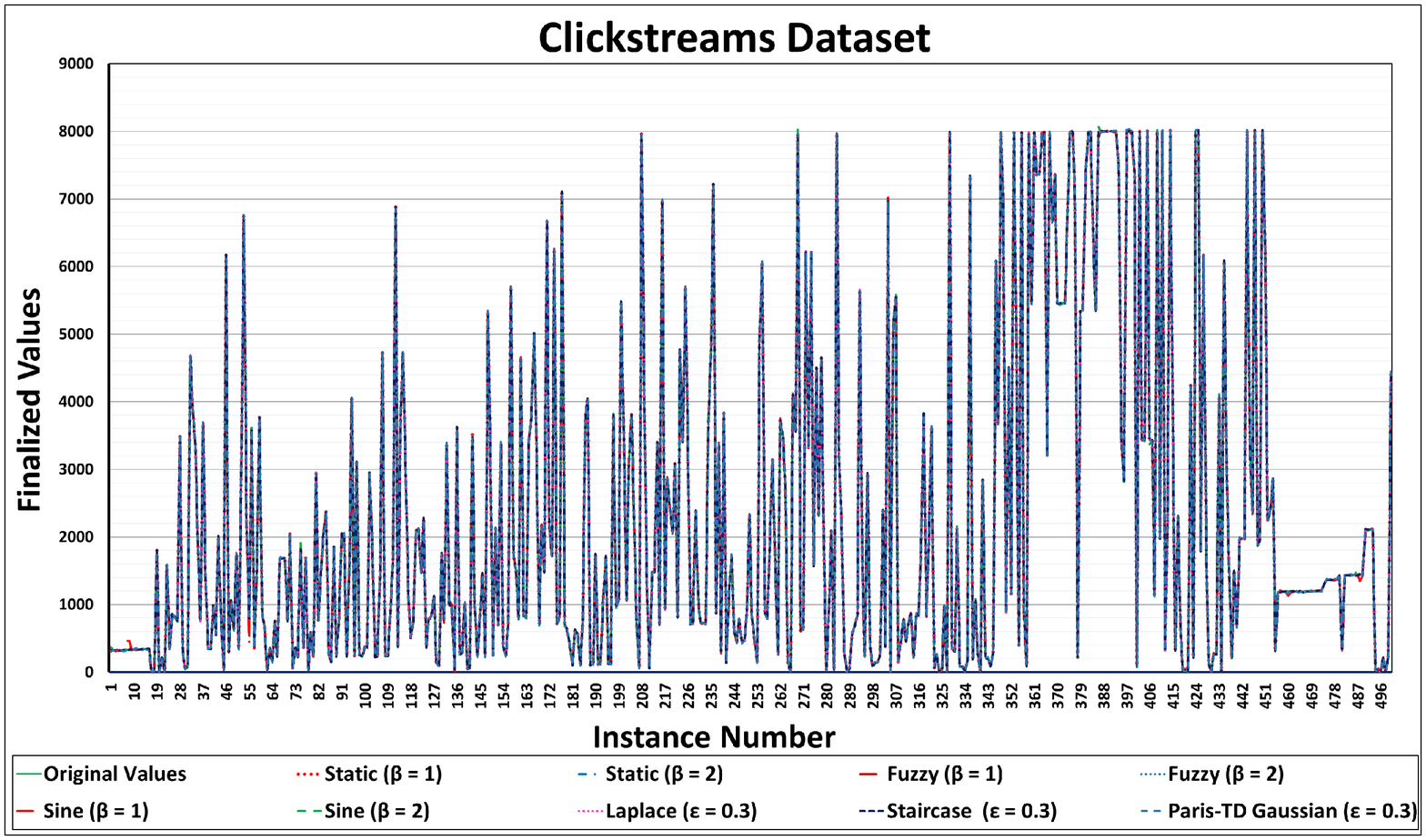}
}
\subfigure[]{
\includegraphics[scale = 0.33]{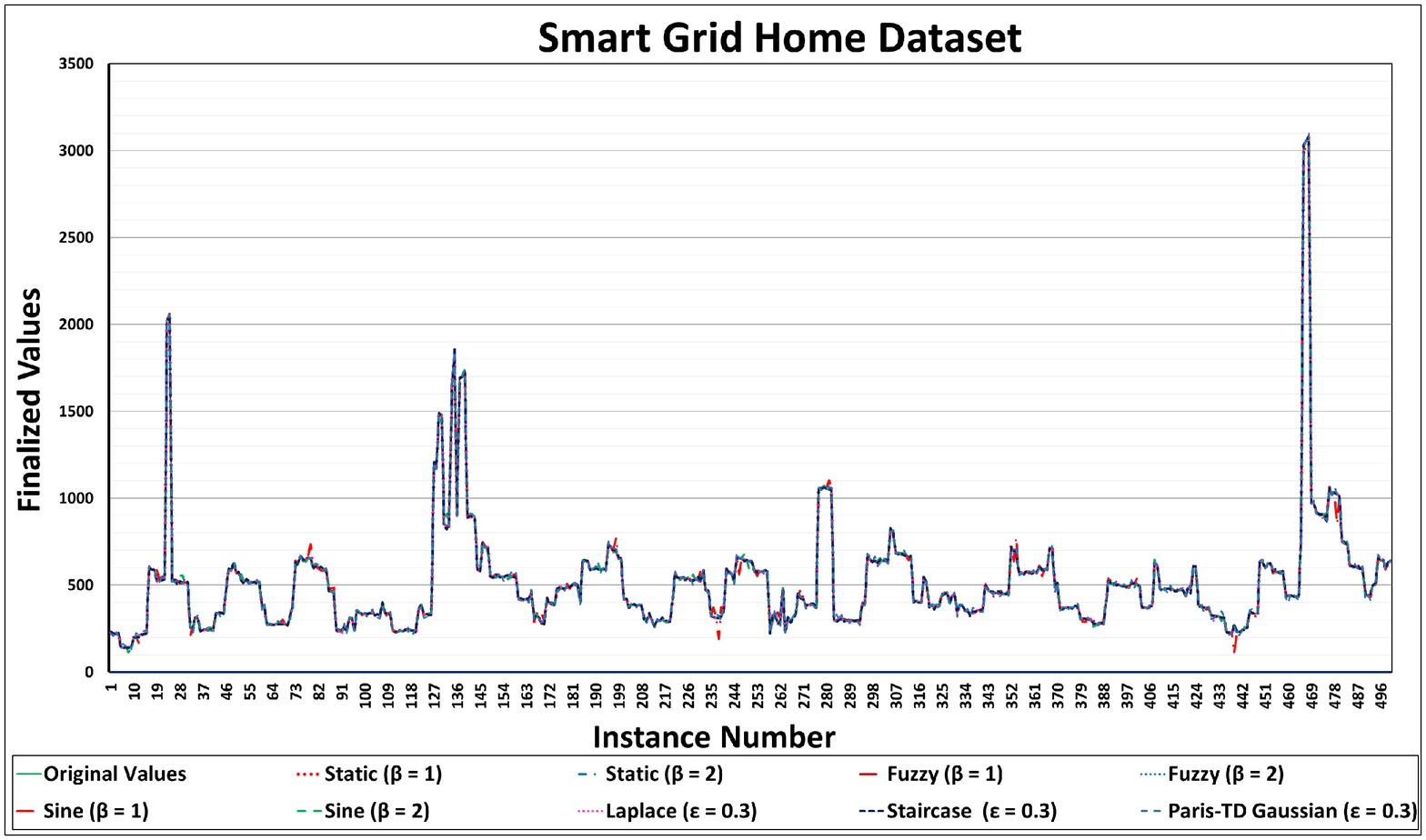}
}
\subfigure[]{
\includegraphics[scale = 0.33]{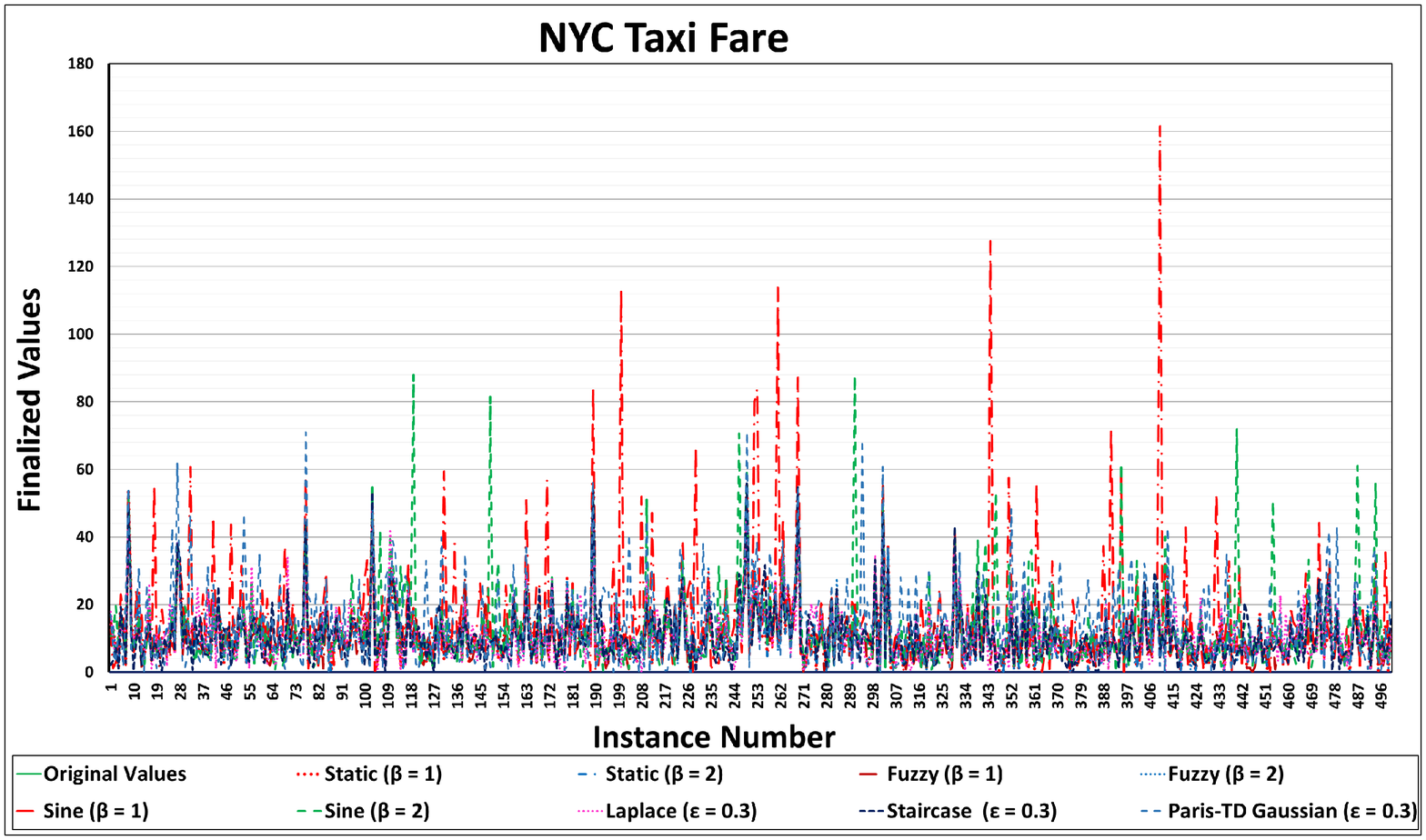}
}

\end{center}
\caption{\textsc{Real-Time Perturbation} of multiple Budget Selection Mechanisms of Huff-DP in comparison with Laplace~\cite{DPRef02}, Paris-TD (Gaussian)~\cite{compareGauss}, and Staircase mechanism~\cite{comparestair} on Real-Time Datasets (a) Flight Distance (b) Clickstreams Dataset (c) Smart Home Energy Usage (d) NYC Taxi Fare Dataset }
\label{fig:Realgraphs}
\end{figure*}



\begin{figure*}[t]
\centering
\captionsetup{justification=centering}
\begin{center}

\subfigure[]{
\includegraphics[scale = 0.75]{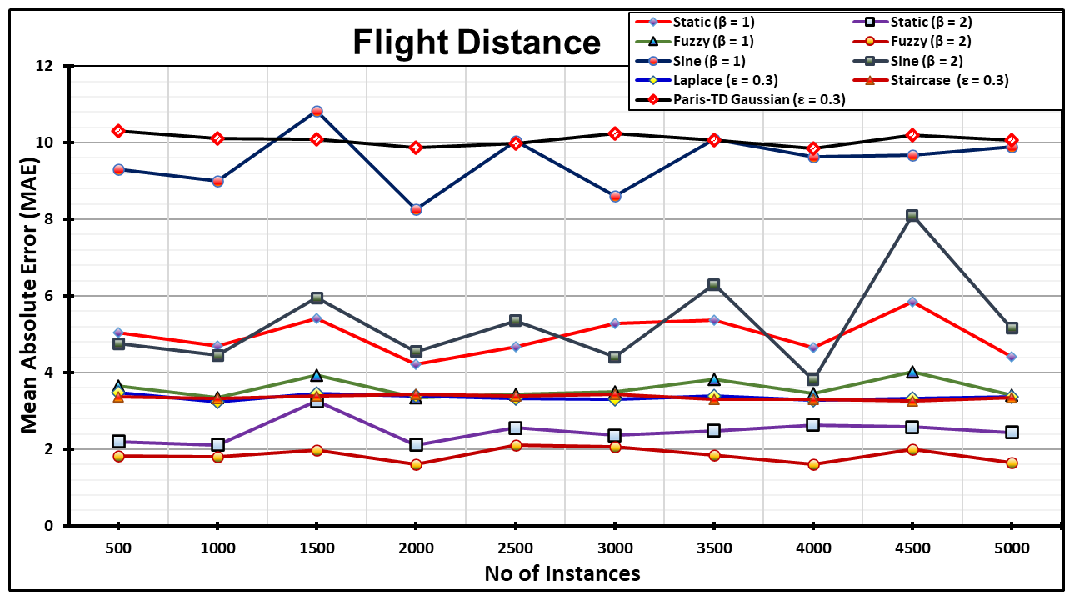}
}
\subfigure[]{
\includegraphics[scale = 0.75]{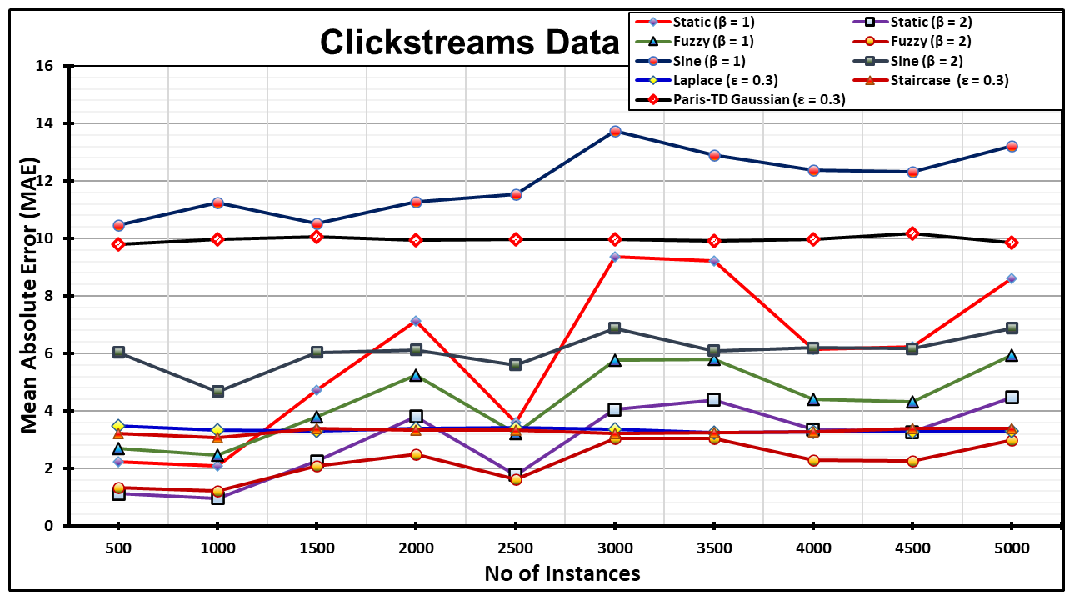}
}
\subfigure[]{
\includegraphics[scale = 0.75]{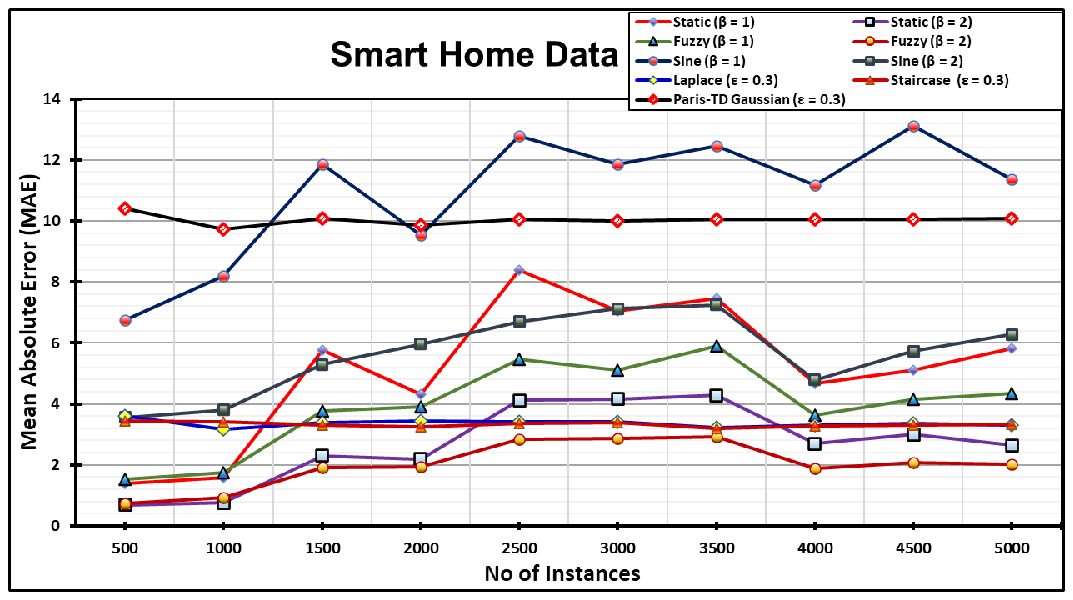}
}
\subfigure[]{
\includegraphics[scale = 0.75]{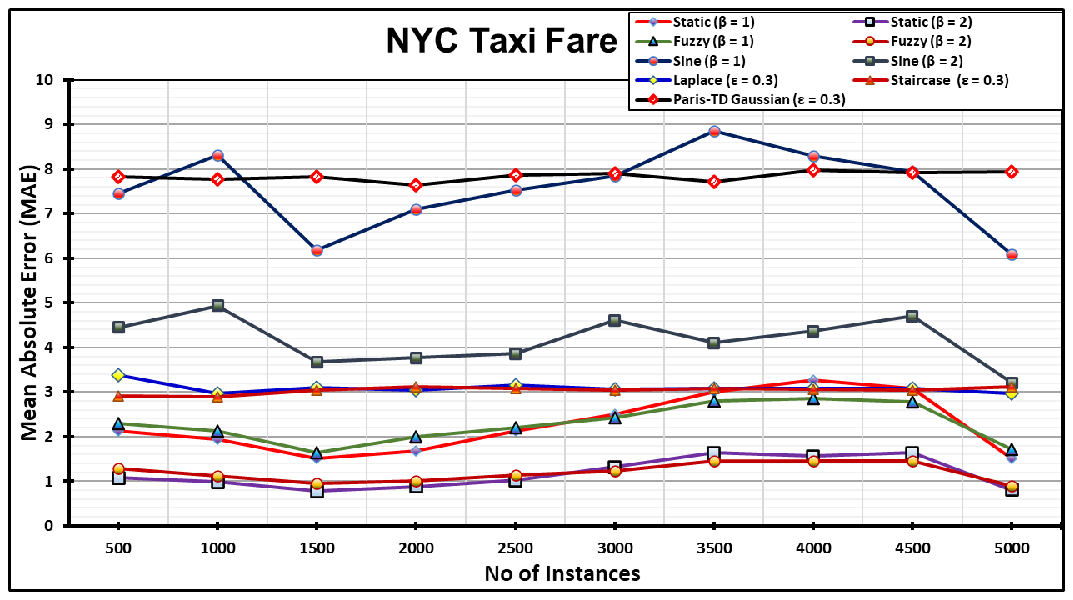}
}

\end{center}
\caption{\textsc{Mean Absolute Error (MAE)} of multiple Budget Selection Mechanisms of Huff-DP in comparison with Laplace~\cite{DPRef02}, Paris-TD (Gaussian)~\cite{compareGauss}, and Staircase mechanism~\cite{comparestair} on Real-Time Datasets (a) Flight Distance (b) Clickstreams Dataset (c) Smart Home Energy Usage (d) NYC Taxi Fare Dataset}
\label{fig:maegraphs}
\end{figure*}


\begin{figure}[t]        
\centering
\includegraphics[scale = 0.8]{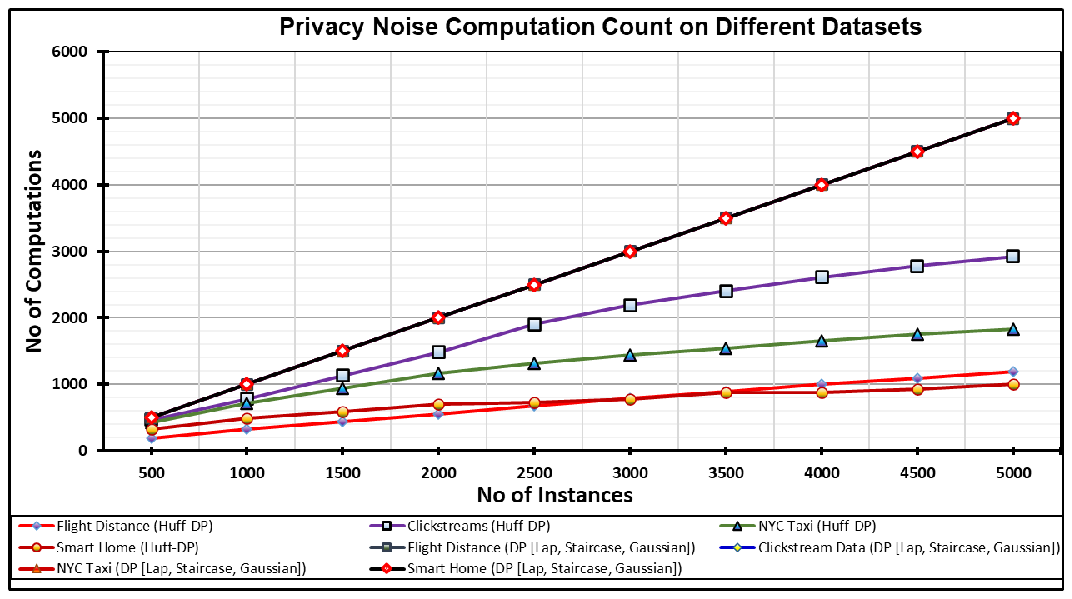}
 \caption{\textsc{No of Computations} required w.r.t Input Instances of Huff-DP in comparison with Traditional DP Notions [Laplace~\cite{DPRef02}, Paris-TD (Gaussian)~\cite{compareGauss}, and Staircase mechanism~\cite{comparestair}] }
  \label{fig:noisecomp}   
\end{figure}

\section{Performance Evaluation of DRDP}\label{PerfSect}
In this section, we describe the performance evaluation and comparison of our Huff-DP mechanism on multiple real-time datasets. 

\subsection{Experimental Settings \& Datasets}
The experiments of our Huff-DP mechanisms are carried out on Python 3.9.12 via Spyder environment, and multiple libraries, such as NumPy, Pandas, HeapQ, and Math. Similarly, from the data perspective, we extract up to 5,000 arbitrary real-time values from the selected datasets. Moreover, we compare our experimental results with the notions of Laplace~\cite{DPRef02}, Parid-TD (Gaussian)~\cite{compareGauss}, and Staircase Laplace mechanism~\cite{comparestair} in order to show the effectiveness from the viewpoint of mean absolute error (MAE), real-time perturbation, and budget computation cost.

\paragraph{Datasets}
For our Huff-DP evaluation, we took four real-world datasets, which are discussed as follows:

\begin{itemize}
\item \textbf{Flight Distance}~\cite{dataair01}: A real-time customer satisfaction scores dataset containing multiple attributes. We used the distance travelled by each customer from the given dataset.
\item \textbf{Kosarak Clickstreams}~\cite{clicksdata01}: A clickstreams online dataset from a Hungarian website containing the clicks patterns. We consider this data as a click over different values categorie. 
\item \textbf{NYC Taxi Fare }~\cite{nyctaxidata01}: A dataset of New York City taxi fares for multiple taxi trips. 
\item \textbf{Smart Home Energy Usage}~\cite{smarthomedata01}: A dataset of smart home with respect to multiple appliances. We use aggregate statistics for each time instance.
\end{itemize}

\subsection{Optimized Private Data Release}

One of the key factors in releasing real-time almost periodic data is to ensure the utility of data, which basically guarantees that the data is still useful for statistical analysis purposes. Finalized values after perturbing with multiple budget selection models of Huff-DP mechanism have been provided in Fig.~\ref{fig:Realgraphs}. By visualizing the given figures, it can be seen that the final released values do not have much difference with respect to the original values. For instance, in Fig.~\ref{fig:Realgraphs}(a), the values of flight distance for travellers has been presented. It can be seen that the noise is a bit higher in case of unique instances, such as during the instance 50-60 or 110-120. Especially in the case of the `sine’ budget selection mechanism, the perturbation is a bit higher, because sine provides a higher perturbation range as compared to others. Moving further to Fig.~\ref{fig:Realgraphs}(b), the dataset of clicks stream can be visualized, where changes can be analyzed from values 1-10, 40-50, etc. Furthermore, in Fig.~\ref{fig:Realgraphs} (c), the real-time smart home energy usage has been presented. Minor modifications from the viewpoint of perturbation can be seen. However, since the data range is pretty high, the perturbation is not much visible. Finally, in Fig.~\ref{fig:Realgraphs} (d), the perturbation values can be clearly seen because the data range is low. It can be seen that for the unique set of values, the sine function provides a strong perturbation. Similarly, after sine, the static model and fuzzy models provide perturbation for required instances. As compared to constant perturbation from Laplace~\cite{DPRef02}, Paris-TD (Gaussian)~\cite{compareGauss}, and Staircase mechanism~\cite{comparestair}, our Huff-DP mechanism only perturbs with higher noise when the value is unique and occurs less frequent. However, for more frequent values, the perturbation is pretty less.\\
It is important to mention that we carry out experiments on up to 5,000 instances. But in the given figures (Fig.~\ref{fig:Realgraphs}), we only present the values of the first 500 instances, as visualizing the changes in 5,000 real-time will not be possible in graphical form. Still, for a number of instances, the perturbation is very minute and hardly noticeable. Similarly, for 500 instances, the repetition among data is not much higher and a number of values are considered as unique values. Thus, perturbation is a bit higher as compared to a higher number of instances. However, this trend changes when we progress towards a higher value range. In order to demonstrate the core error difference, we also evaluate MAE, which is presented in the next section.

\subsection{Mean Absolute Error (MAE)}
One of the key parameters while evaluating differential privacy perturbation results is MAE. We carry out performance evaluation of our multiple budget selection mechanisms of Huff-DP algorithm on real-time datasets with respect to MAE. The formula to calculate MAE is given as follows~\cite{maeref01}:
\begin{equation}
MAE = \frac{1}{N_{O_r}}\sum_{j=1}^{N_{O_r}}|C_{R} – P_{R_v}|
\end{equation}
The graphical illustration of MAE with respect to various real-time datasets and multiple privacy mechanisms is given in Fig.~\ref{fig:maegraphs}. In Fig.~\ref{fig:maegraphs}(a), the values of MAE for flight distance data can be visualized. E.g., throughout the graph, the value of fuzzy logic based budget selection mechanism (with $\beta = 2$) outperforms all other mechanisms by providing minimum MAE. Afterwards, the static budget mechanism (with \$beta = 2) outperforms the remaining ones. After that, a combination of multiple mechanisms can be seen accordingly. Similarly, the mechanism based upon sine budget selection provides a higher level of privacy because it introduces maximum randomness in budget selection, that is why it provides a high perturbation range. The same trend can be seen in case of clickstream data in Fig.~\ref{fig:maegraphs}(b), where the value of MAE is pretty minimum for majority of budget selection functions especially when the hyper parameter $\beta = 2$ is selected for computation as compared to Laplace~\cite{DPRef02}, Paris-TD (Gaussian)~\cite{compareGauss}, and Staircase mechanism~\cite{comparestair}. Moving further to Fig.~\ref{fig:maegraphs} (c) and (d), a similar trend in MAE can be seen that fuzzy logic based budget selection outperforms others, and after fuzzy logic, static budget selection provides better results. \\
Overall, it will not be wrong to say that fuzzy logic based, and static function based budget selection mechanisms outperforms because of the reason that they provide a better control over the selection of privacy budget, as compared to sine function based budget selection. For instance, in case of Level 1, the budget will be selected from a higher range in case of fuzzy logic and static. But even in the case of level 1 with sine function, the budget can range from 0.01 – 1. Thus, for an overall high perturbation, sine can be chosen. However, for a dynamic range of perturbation, fuzzy logic based budget selection outperforms others.

\begin{table}[ht!]
\begin{center}
 \centering
 \scriptsize
 \captionsetup{labelsep=space}
 \captionsetup{justification=centering}
 \caption{Comparative Analysis of Huff-DP with other Differential Privacy Models.}
  \label{tab:comparehuff}

\begin{tabular}{|p{7em}|p{4em}|p{4em}|p{4em}|p{4em}|}

\hline

\textbf{Mechanism} & \textbf{Dynamic Budget Allocation} & \textbf{Protecting Unique Instances} & \textbf{Frequency based Privacy} & \textbf{Huffman Compression} \\ \hline

\textbf{Huff-DP (Proposed)} & Yes & Yes & Yes & Yes   \\ \hline

\textbf{Laplace~\cite{DPRef02}} & No & Partial & No & No   \\ \hline

\textbf{Paris-TD (Gaussian)~\cite{compareGauss}} & No & Partial & No & No   \\ \hline

\textbf{Staircase Laplace~\cite{comparestair}} & No & Partial & No & No   \\ \hline

 \end{tabular}
  \end{center}
\end{table}


\subsection{Optimal Privacy Noise Computational Count}

Since, we are calculating the privacy budget for each value with respect to its occurrence frequency. Therefore, even if a specific character/value/data occurs more than once, we still have to calculate the privacy budget and noise value only once, because we have determined it with respect to privacy level. A graphical illustration of computational count of differential privacy noise via Huff-DP mechanism in comparison with other conventional mechanisms has been provided in Fig.~\ref{fig:noisecomp}. It can be seen that Huff-DP mechanism outperforms in all selected datasets which have repetitive real-time values in an almost periodic manner. For example, the datasets of flight distance and smart home have more repetitive values, thus the computational count cost of Huff-DP is minimal. E.g., in case of 2,500 instances, the traditional differential privacy models have to compute noise for 2,500 instances, however, in case of Huff-DP, it ranges from 800 – 2,000 depending upon the data and repetition.

\subsection{Discussion}
By analysing the experimental results of perturbed values, MAE, and noise computation count, it can be said that our proposed Huff-DP mechanism outperforms other mechanisms in terms of providing a dynamic budget selection mechanism with respect to data frequency on the basis of Huffman-tree depth. Furthermore, from within the Huff-DP mechanism, we propose three budget allocation mechanisms named as static, sine, and fuzzy logic based selection. Among these models, it can be said that if an application scenario requires much higher privacy even for early levels, then sine mechanism is the best choice. Contrarily, if utility is the first aim and high and controlled perturbation is required majorly for unique values, then fuzzy logic based, and static selection based mechanisms can be optimal choice.

\section{Conclusion}

The evolution of communication technologies and interconnected devices resulted in generation of a huge amount of real-time data, which is being used to carry out a number of statistical and analytical tasks and predictions). Nevertheless, this data is helpful in many ways, but strong models are required in order to store and transmit it optimally from a compression and privacy perspective. In order to store and disseminate this data, source coding based models are usually used (such as Huffman coding, etc). Similarly, in order to protect privacy, differential privacy is usually used as a de facto mechanism. Nevertheless, differential privacy provides good privacy guarantees, but it needs further amendments with respect to real-time data (e.g., frequency based data). Thus, in this paper, we work over integrating the notion of Huffman coding with differential privacy and proposed Huff-DP mechanism, which works over the phenomenon of perturbing the values with respect to their privacy level, which is derived with the help of Huffman tree. We further propose three sub-algorithms of Huff-DP model, named as static, sine-based, and fuzzy logic based budget selection model, which selects optimal privacy budget with respect to privacy requirements. Finally, we carry out a performance evaluation of our proposed model and the experimental results demonstrate that our work outperforms other traditional differential privacy mechanisms in terms of MAE and noise computational count.

\bibliographystyle{IEEEtran}


\end{document}